\documentclass[11pt]{article}
\usepackage{setspace}
\usepackage[utf8]{inputenc}
\usepackage{authblk}
\usepackage[hidelinks]{hyperref}
\usepackage{graphicx}
\usepackage{amsmath}

\begin{document}
	
	\title{Grounds for trust: Essential Epistemic Opacity and Computational Reliabilism}


\author{Juan M. Dur\'an\footnote{Delft University of Technology} - Nico Formanek\footnote{High Performance Computing Center Stuttgart - University of Stuttgart}\\\textit{Minds and Machines} - (2018) 28 (4):645–666\\ DOI:10.1007/s11023-018-9481-6}


%
\date{}

\maketitle

	\begin{abstract}
		Several philosophical issues in connection with computer simulations rely on the assumption that results of simulations are trustworthy. Examples of these include the debate on the experimental role of computer simulations \cite{Parker2009, Morrison2009}, the nature of computer data \cite{Barberousse2013, Humphreys2013}, and the explanatory power of computer simulations \cite{Krohs2008, Duran2017}. The aim of this article is to show that these authors are right in assuming that results of computer simulations are to be trusted when computer simulations are reliable processes. After a short reconstruction of the problem of \textit{epistemic opacity}, the article elaborates extensively on \textit{computational reliabilism}, a specified form of process reliabilism with computer simulations located at the center. The article ends with a discussion of four sources for computational reliabilism, namely, verification and validation, robustness analysis for computer simulations, a history of (un)successful implementations, and the role of expert knowledge in simulations.
	\end{abstract}

\section{Introduction}

In a recent dispute over the philosophical novelty of computer simulations, Paul Humphreys \cite{Humphreys2009} argued in favor of four genuine philosophical issues that require the attention of philosophers, namely, \textit{epistemic opacity}, the \textit{semantics} of computer simulations, the \textit{temporal dynamics} of computational processes and the distinction \textit{in principle/in practice}. Of those four issues, this article focuses solely on \textit{epistemic opacity}, arguably the most controversial issue raised by Humphreys. There are at least two good reasons for paying attention to epistemic opacity. First, because it is the most direct consequence of the so-called \textit{anthropocentric predicament}, that is, the claim that humans have been displaced from the center of production of knowledge. Thus understood, any epistemological treatment involving computer simulations faces the question of epistemic opacity. Second, because Humphreys did not offer any suggestions for a solution to epistemic opacity, but rather restricts his analysis to pointing it out as a philosophically novel issue. In this context, several questions remain unanswered. For instance, `how does epistemic opacity affect the epistemological treatment of computer simulations?', `in the context of opacity, could it be correct to say that scientists are overemphasizing the success of simulations?' \cite{Frigg2009}, and `is there a way to conceive of epistemic opacity coexisting with some form of knowledge?' This article aims at addressing these questions and providing a qualitative answer to epistemic opacity in the context of computer simulations. Furthermore, it also offers a formal framework to secure claims about knowledge provided by computer simulations.

To frame the issue within the current philosophical debate, consider the question of whether epistemic opacity is an unavoidable issue in scientific practice, and thus is the acceptance of the epistemic superiority of computers. This point has been recently criticized by Julian Newman, who argues that epistemic opacity is a non starter for the epistemological treatment of computer simulations. To him, epistemic opacity is a symptom of modelers having failed to adopt sound practices of software engineering \cite{Newman2015}. Instead, by means of developing the right engineering and social practices, Newman argues, modelers would be able to avoid several forms of epistemic opacity and ultimately reject the assertion that computers are a superior epistemic authority.\footnote{Let us note that Newman takes epistemic opacity as a condition for the anthropocentric predicament. Humphreys, instead, takes epistemic opacity as a consequence of the anthropocentric predicament. Both interpretations are possible and both find an answer in this article.} As he explicitly puts it: ``[...] well architected software is not epistemically opaque: its modular structure will facilitate reduction of initial errors, recognition and correction of those errors that are perpetrated, and later systematic integration of new software components'' \cite[267]{Newman2015}.\footnote{Many philosophers have engaged with the problem of epistemic opacity and suggested reliabilism as the most suitable solution, although most of them have not provided a full fledged account. A shortlist includes \cite{Lenhard2010}, \cite{Hasse2017} and \cite{Kaminski2017}, among others. The exception is \cite{Duran2014}, who offers an early attempt to reconstruct reliabilism in the context of computer simulations.}

Although we find Newman's concerns reasonable, for they are based on the assumption that knowing how a method works gives insight into its outcome, we do not agree with his conclusion. Software engineering practices also promote genuine forms of epistemic opacity. For instance, Timothy Colburn and Gary Shute claim for new forms of abstraction exclusively for computer systems that hide but do not neglect specific aspects of the target system. That is, standard forms of \textit{abstraction}, \textit{idealization} and the like aim at neglecting specific aspects of the target system, while \textit{information hiding} consists in hiding ``details that are essential in a lower-level processing context but inessential in a software design and programming context'' \cite[176]{Colburn2007}. Understanding these claims in our context, it is possible to identify unavoidable degrees of opacity in standard software engineering practice that come with an agent being unable to relate a given computer program with its physical instantiation on the computer machine (i.e., information is hidden to the agent).

A more general solution is therefore required, one that allows researchers to acknowledge epistemic opacity but not at the expense of losing knowledge. This viewpoint is also shared by other philosophers, such as John Symons and Jack Horner who argue that although it is impossible to test the correctness of all possible paths of computer software within any human timescale, trusting the results of computer software is nevertheless possible \cite{Symons2014}. This article provides precisely such a general solution.

To be more precise, this article offers an alternative analysis to the current philosophical treatment on epistemic opacity. In here, a formal framework is developed to the effect of allowing knowledge provided by computer simulations without rejecting some degrees of epistemic opacity. To this end, the article first analyzes \textit{epistemic opacity} and \textit{essential epistemic opacity} as presented by \cite{Humphreys2009}. This is the main subject of section \ref{EO}, where epistemic opacity is reconstructed in terms of accessibility and surveyability conditions on justification. Section \ref{dissolving} elaborates on \textit{computational reliabilism}, a version of \textit{process reliabilism} \cite{Goldman1979} more suitable for accommodating computer simulations. The aim of this section is to understand the implications of epistemic opacity for the analysis of knowledge. It is worth noting that although this article focuses efforts on computer simulations, much of what is said here can also be extended to other uses of computers provided that the right methods for grounding computational reliabilism are in place. Section \ref{four_sources} further elaborates on computational reliabilism by advancing the four sources that attribute reliability to computer simulations. The last section recapitulates the findings and advocates for further issues of genuine philosophical interest.

\section{Epistemic Opacity (EO)} \label{EO}

In the following, we reconstruct epistemic opacity in terms of accessibility and surveyability conditions on justification. 

\subsection{What is epistemic opacity?}

Before we give formal definitions of epistemic opacity, we note that it is motivated by a sceptical line of thought. This is best seen by considering the following example: take the uncontroversial assumption that human intellect is limited in the sense that it cannot be acquainted with every natural number -- for there are an infinity of them. It is a well known fact that much of modern mathematical reasoning depends upon properties of infinite sets (e.g. the statement that every composite number has a prime factorization). The skeptical challenge is now this: how do mathematicians establish the truth of general statements which obviously transcend their intellect? They certainly cannot try out every instance. One way to answer this question is to employ specific methods such as mathematical induction and proof by contradiction for conferring the required certainty. The question that naturally follows is what warrants the validity of those latter methods?

This question concerned early 20th century mathematicians and philosophers in the foundational crisis of mathematics, and has its analogue in the notion of epistemic opacity that interests us now. While mathematicians sought to justify their use of infinitary methods, which were at the core of Hilbert's programm (see \cite{Zach2016} for an overview), we seek to justify computational methods such as computer simulations. Such computational methods exceed human abilities in a similar way because they also involve a large number of steps to be acquainted with. 

As mentioned earlier, Humphreys has stated that a general account of justification -- or a full fledged epistemology -- for computer systems has to be non-anthropocentric, that is, computers replace humans in the process of justification. The formal definition of epistemic opacity proposed by Humphreys in \cite{Humphreys2004} and repeated in \cite{Humphreys2009} goes as follows:

\begin{quote}
	[A] process is epistemically opaque relative to a cognitive agent X at time t just in case X does not know at t all of the epistemically relevant elements of the process \cite[618]{Humphreys2009}
\end{quote}

Concerning Hilbert's program, it has been debated which methods are finitary and therefore admissible. In the context of computer systems, and more specifically to our interests, computer simulations, the question is, which are the exact limitations of the cognitive agent in Humphrey's definition. It is clear that this has to do with what the agent can and can not know. Now, since \emph{can} is a modal verb, we will give an interpretation of the modality involved; furthermore, we will also give an account of the epistemic \emph{know}. 

\subsection{Reconstruction}

Our reconstruction focuses on ``X does not know'' as part of the above definition, and analyzes \emph{knowing} in terms of accessibility and surveyability. The central role of justification in the computing process will be argued for. We note in passing that \emph{knowing} could also have been analyzed in terms of JTB. As it is contentious that JTB gives sufficient conditions for knowledge one would have to resort to a more general reading like JTB+X. While the role of justification becomes directly apparent in JTB+X due to the justification clause, we think that avoiding the immediate replacement \emph{know $\rightarrow$ JTB+X} clauses is more charitable to Humphrey's intentions and thus allowing for a less contentious interpretation of knowledge.

As we focus on the justificatory aspects of epistemic opacity, we should clarify what we mean by justification. A simple example could be a (not necessarily deductive) argument as exemplified in ones (not necessarily deductive) logic of choice. An argument has premises, steps and a conclusion, with the inference from the premises to the conclusion warranted by some inferential relation. We also allow for informal and non-deductive arguments, so justification doesn't have to be truth-preserving nor do we require a formal theory of truth. Though it seems obvious that computational processes are deductive \cite{Beisbart2012}, we don't require them to be formalized in some specific logic nor will we attempt to do so.

Let us now begin by emphasizing the last part in Humphreys' definition of EO:

\begin{quote}
	[A] process is epistemically opaque relative to a cognitive agent X at time t just in case X does not know at t all of \emph{the epistemically relevant elements of the process} \cite[618]{Humphreys2009}
\end{quote}

A process which contains epistemically relevant elements is different from a causal process or a mere sequence of events. We take this to mean that such a process can be understood as a justification, which is an epistemic term in itself. Basically, this means that the epistemically relevant elements of the process are the steps of the argument giving the justification.

In the next step we deal with the \emph{knowing} part of the definition, while already substituting \emph{steps of the justification} for \emph{epistemically relevant elements}.

\begin{quote}
	[A] process is epistemically opaque relative to a cognitive agent X at time t just in case X does \emph{not know} at t all of the steps of the justification
\end{quote}

Taking into account the speed and volume at which justificatory steps could be generated by computing processes, there are two conditions for how an agent can fail to know them. She could fail to access each step -- maybe they are generated and discarded so fast that she can not keep up, maybe her memory is insufficient. But even granting full access to every justificatory step, will she still be able to check every step according to some predefined set of rules? In most cases the answer is probably no, since having access to the steps of a process and checking the validity of an argument are two different things. All things being equal, the limiting factors to be able to check a justification are finiteness in length and time. This is exactly what the concept of \emph{surveyablity} captures for mathematical proofs \cite{Tymoczko1981}. Following this idea, we say that \emph{a justification is surveyable if and only if it is finite in length and checkable in finite time}. Of course the exact amount of finitude and the meaning of checkability depends on the agent. We will discuss the application to human agents in the next section. 

With these clarifications in mind, our final proposal for the definition of epistemic opacity including the notions of accessibility and surveyability now reads:

\begin{quote}
	[A] process is epistemically opaque relative to a cognitive agent X at time t just in case X at t doesn't have access to and can't survey all of the steps of the justification.
\end{quote}

We now see that a justification is epistemically opaque if there is a failure to justify elements of it. At this point an example of such a computational process is in order. Consider the execution of a computer program on a particular machine (i.e. the changes in the physical state of the processor, the access to ram memory, etc.). The steps involved in the justification are defined relative to an agent. That is, if one already knows how a half-adder works, one is allowed to subsume its internal processes and take the result as justified. Say you know the workings of the basic logical circuitery of the CPU and the rest of the hardware. You are now going to write a `hello world' in machine code. Would the execution of this program be epistemically opaque to you? The answer is no, and the reason is because you \emph{can} give a justification for every step thanks to your previous knowledge (i.e., of knowing the basic logical circuitery of the CPU and the rest of the hardware). Naturally, it is easy to imagine programs more complex than a simple `hello world' on the screen where giving an actual justification would take too long. The next section tries to capture this aspect of computer software.

\subsection{Strengthening EO with Essential Epistemic Opacity} \label{EEO}

Essential Epistemic Opacity (EEO) imposes severe restrictions on what counts as a justification under EO, depending on the reading of (im)possibility in the following definition:

\begin{quote}
	A process is essentially epistemically opaque to X if and only if it is \emph{impossible}, given the nature of X, for X to know all of the epistemically relevant elements of the process \cite[618]{Humphreys2009}
\end{quote}

\noindent Translating this definition into our scheme from above, it reads:

\begin{quote}
	A process is essentially epistemically opaque to X if and only if it is \emph{impossible}, given the nature of X, for X to \emph{have access to and be able to survey} all of the relevant elements of the justification.
\end{quote}

For Humphreys EEO is on the practical side of his \emph{in principle} - \emph{in practice} divide. For him it makes a difference whether the agent knows something or whether he only \emph{can know} something. Of course his interpretation of the \emph{can know} has to involve a specific account of modality in which human limitations come to play.

The standard account in epistemology interprets \textit{possibilities} as \textit{logical possibilities}, meaning that everything that does not imply a contradiction is therefore possible. This idealizing account of possibilities also affects the nature of X, insofar as the nature of X should not contain anything which prevents logical possibilities to be known. Therefore, on the face of it, the ability of X to access and survey all of the relevant elements, even given its nature, does not imply a logical contradiction.

To make sense of EEO, therefore, we need to further restrict the notion of possibility to cover only \textit{metaphysical} or \textit{physical possibilities}. For the metaphysical case, one needs an account for X's nature or essence which has been contested at least since Aristotle presented the idea. In the end, the extension of EEO would have to be set by a dogmatic decree. Now, if we want to restrict ourselves to physical possiblities,\footnote{An anonymous reviewer pointed out that those possibilities do not depend on the physical theory we adopt. Although correct, we also do not have access to physical possiblities outside of any physical theory. If we want to give meaning to epistimic opacity through physical possibilities, then we have to adopt some physical theory.} then we have to be precise about which physical theory is at stake, for Newtonian possiblities are certainly different from relativistic possibilities.

We propose a common sensical reading of possiblity. For Humphreys, eligible possibilities are restricted by human action\footnote{Humphreys says: ``For certain philosophical purposes, such as demonstrating that some kinds of knowledge is impossible even in principle, in principle arguments are fine. But just as humans cannot in principle see atoms, neither can humans in principle.'' \cite[624]{Humphreys2009}. Let it be noted that we have reproduced the quote verbatim. However, we believe Humphreys meant to say ``For certain philosophical purposes, such as demonstrating that some kinds of knowledge \textit{are} impossible even in principle, in principle arguments are fine. But just as humans cannot in \textit{practice} see atoms, neither can humans in \textit{principle} [or vice versa].''} which takes place in finite time and space and is constrained by finite mental capabilities. A process is EEO if it is impossible for someone to justify epistemically relevant elements of a process under those conditions. Therefore, justifications themselves are restricted by finite space, time and mental capabilities. This means that unsurveyablility is a real possibility for such justifications. In a sense, this claim was already contained in our reformulation of EO, with the difference that the agent referred to was a generic one while now it is a human agent. And it is for humans that most processes in (computational) science are EEO (see \cite{Humphreys2009}). This immediately implies the skeptical challenge mentioned at the beginning, namely, that if most processes are EEO then we are not justified in believing their results.

This article holds a different position. At its core, we claim that researchers are justified in believing the results of computational processes, such as computer simulations, given certain conditions for their reliability. In the following sections, we flesh out these ideas by introducing \textit{computational reliabilism}, a version of \textit{process reliabilism} that relaxes the demands for accessibility and surveyability of the computational process and thereby allows us to regain some human control over justifications.  

\section{Dissolving EEO with computational reliabilism} \label{dissolving}

As mentioned earlier, a core epistemological concern in studies on computer simulation is to find grounds for claims about knowledge. That is to say, to be able to justify the belief that either the results of computer simulations are correct of the target system, that they are valid with respect to our system of beliefs, or simply that they are employed within their intended uses. This is a concern that can be found, explicitly or implicitly, in the work of most philosophers interested in the epistemological input of computer simulations. Margaret Morrison, for instance, explicitly addresses these concerns in her treatment of the `materiality' of computer simulations \cite{Morrison2009}, and reuses the same ideas in her analysis of the role of computer simulations in finding the Higgs boson \cite{Morrison2015}. Wendy Parker is another philosopher that has made explicit her concerns about knowledge and justification in the context of computer simulations. To her mind, conclusions about the target systems on the basis of computer simulation results ``cannot be automatic, but rather require justification'' \cite[490]{Parker2009}.

Accepting EEO in computer systems has several consequences of importance for the epistemological treatment of computer simulations. Perhaps the most immediate one is that EEO casts doubts on whether researchers could trust the results of computer simulations at all. But by removing trust in results, the initial reasons for using computer simulations in scientific and engineering practice falter. Furthermore, standard philosophical examination on the experimental value of computer simulations becomes meaningless \cite{Morgan2003, Morgan2005}. Without the assumption that results of computer simulations are trustworthy, there are no grounds for claims about their experimental side.

As discussed, a core assumption lying behind EEO is that the justification of results requires some form of surveyability of the simulation process. That is to say, in order to be justified in believing the results of the simulations, researchers must survey every step of the computational process that leads to such results. But as it has been argued earlier, such surveillance is indeed impossible and, for our purposes, equally undesirable. Therefore, a more effective solution needs to be found, one that allows some degree of reliability to be attributed to the simulation process and, by means of it, to grant the results the necessary trust. Now, since EEO prevents us from attributing reliability by means of inspecting the simulation models or by surveying the process of computing such models, then it must be done by appealing to procedures external to the simulation itself. 

In this context, many philosophers have suggested different sources for reliability. Claims range from stating that good simulations require well grounded scientific knowledge \cite{Massimi2015}, to assertions that scientists believe the results of their simulations because they trust the assumptions upon which they are built \cite{Beisbart2017}. Although we heartedly endorse these claims, more needs to be said. Relating computer simulations to well grounded scientific knowledge as well as trusting the assumptions built in are, at best, only necessary but not sufficient conditions for attributing reliability to computer simulations.  For this reason, we argue that these sources are more diverse and numerous than those usually discussed in the literature. In this respect, we offer the first comprehensive review of the sources that attribute reliability to computer simulations and, by doing so, grant trust to their results. Furthermore, claims about knowledge need to be located within a theoretical framework that properly articulates these sources and supplies a justification of the reliability of the computer simulation along with reasons to believe in their results. To us, such framework comes in the form of a modified version of Alvin Goldman's \cite{Goldman1979} \textit{process reliabilism} -- or reliabilism for short -- that we deem to call \textit{computational reliabilism}. Let us now try to make sense of these ideas.

\subsection{Computational reliabilism}


Let us begin by presenting \textit{process reliabilism} as elaborated by \cite{Goldman1979} and \cite{Goldman2016}. In its simplest form, reliabilism can be expressed in the following way:\\

\begin{quote}
	(PR) if \textit{S}'s believing \textit{p} at \textit{t} results from \textit{m}, then \textit{S}'s belief in \textit{p} at \textit{t} is justified.\\
	
	where \textit{S} is a cognitive agent, \textit{p} is any truth-valued proposition, \textit{t} is any given time, and \textit{m} is a reliable process.
\end{quote}

Thus understood, according to reliabilism, Peter is justified in believing that `2 + 2 = 4' because counting small natural numbers is usually a reliable process. Indeed, there is nothing accidental about the truth of the belief that `2 + 2 = 4' when knowledge is acquired by a reliable reasoning process such as doing arithmetics under normal circumstances and within a limited set of operations. 

For process reliablism to work, however, it is essential that a reliable process is not so because it was successful once, but rather because there is a tendency to produce a high proportion of true beliefs relative to false ones. Goldman has a simple way to depict reliabilism as ``consist[ing] in the tendency of a process to produce beliefs that are true rather than false'' \cite{Goldman1979}. His proposal, then, highlights the place that a belief-forming process has in the steps towards knowledge. Consider the following example offered by Goldman:

\begin{quotation}
	If a good cup of espresso is produced by a reliable espresso machine, and this machine remains at one’s disposal, then the probability that one’s next cup of espresso will be good is greater than the probability that the next cup of espresso will be good given that the first good cup was just luckily produced by an unreliable machine. If a reliable coffee machine produces good espresso for you today and remains at your disposal, it can normally produce a good espresso for you tomorrow. The reliable production of one good cup of espresso may or may not stand in the singular-causation relation to any subsequent good cup of espresso. But the reliable production of a good cup of espresso does raise or enhance the probability of a subsequent good cup of espresso. This probability enhancement is a valuable property to have. \cite[28]{Goldman1979}
\end{quotation}

The probability here is interpreted objectively, that is, as the chance that a recorded observation -- or a long history of collected data -- produces beliefs that are true rather than false. The core idea of reliabilism is that if a given process is reliable in one situation, then it is very likely that, all things being equal, the same process will be reliable in a similar situation. Thus, Peter is justified in believing that `2 + 2 = 4' because he has been correct in the past -- and, we could add, most likely he will also be correct in the future. Let it be noted that Goldman is very cautious in demanding infallibility or absolute certainty for the reliabilist account. Rather, a long-run frequency or propensity account of probability furnishes the idea of a reliable production of coffee that increases the probability of a subsequent good cup of espresso.

Now, one way to reinterpret Goldman's reliabilism in the context of computer simulations is to say that researchers are justified in believing the results of their simulations because there is a reliable process that produces, most of the time, true beliefs about such results. We can now reinterpret computational reliabilism in the following terms:

\begin{quote}
	
	(CR) if \textit{S}'s believing \textit{p} at \textit{t} results from \textit{m}, then \textit{S}'s belief in \textit{p} at \textit{t} is justified.\\
	
	where \textit{S} is a cognitive agent, \textit{p} is any truth-valued proposition related to the results of a computer simulation, \textit{t} is any given time, and \textit{m} is a reliable computer simulation.
\end{quote}

Let us note that the formulation of process reliabilism remains largely unmodified by computational reliabilism, as it is evidenced in (CR). An important -- and rather obvious -- difference, however, is that process reliabilism is no longer a general account for any \textit{p} and \textit{m}, but rather specified for computational undertakings. In this respect, computational reliabilism takes that \textit{p} is a truth-valued proposition related to the results of a computer simulation. These could be particular, such as `the results show that republicans have won,' `the results suggest an increase of temperature in the Arctic as predicted by theory', and `the results are consistent with experimental results,' among others. Alternatively, they could also be general such as `the results are correct of the target system', `the results are valid with respect to the researcher's corpus of knowledge', and `the results are accurate for their intended use.'\footnote{By posing these general propositions, we remain neutral on whether computational reliabilism should hold commitments to a representationalist viewpoint (e.g., first general proposition), or to a non-representationalist one (e.g., second and third general propositions). Furthermore, it is important to notice that we have spelled out computational reliabilism in a positive form, that is, that if researchers know \textit{p} then they `cannot be wrong' about the results. However, and just like process reliabilism, computational reliabilism also makes place for the possibility of errors \cite{Goldman2016}.} Naturally, the reliable process \textit{m} is identified with the computer simulation (see section \ref{justifying_CR} for further differences with process reliabilism).

We can now assimilate Goldman's process realibilism into our analysis of computational reliabilism: researchers are justified in believing the results of their simulations when there is a reliable process (i.e., the computer simulation) that yields, most of the time, trustworthy results. More formally, the probability that the next set of results of a reliable computer simulation is trustworthy is greater than the probability that the next set of results is trustworthy given that the first set was produced by an unreliable process by mere luck \cite{Duran2014}. 

The challenge now is to spell out what makes a computer simulation a reliable process in the sense given above. To this end, section \ref{four_sources} discusses four sources for computational reliabilism. However, let us first discuss some shortcomings of process reliabilism for computer simulations and the reasons for promoting (CR).


\subsection{Justifying computational reliabilism} \label{justifying_CR}



Earlier, we mentioned that accepting EEO casts doubts on whether researchers could trust the results of computer simulations at all. The chief argument for EEO is, again, that computer simulations contain so many steps that they become inaccessible and unsurveyable by a human agent, and thus their belief in the results are impossible to justify. This is a fundamental skeptical concern about knowledge and justification for computer simulations with the consequences already stated. 

Computational reliabilism has been the proposed solution to the skeptical challenge posed by EEO. Taken seriously, however, EEO entails that the processes that attribute reliability to the simulation could be, in turn, also epistemically opaque. This is to say that the alleged reliability of computer simulations could be attributed by an unreliable process. To illustrate this point, consider that researchers cannot trust the results of a simulation using the Schelling's model of segregation if the original distributions are produced by a pseudo-random generator that produces non-random results (i.e., it is a non-reliable pseudo-random generator). Trusting the results of computer simulations, therefore, depends on having a chain of reliable processes that, in the end, allow researchers to be justified in believing the results.\footnote{This point raises the obvious question of where does this chain of reliable processes end. Unfortunately, we cannot answer this question here.}

Unfortunately, process reliabilism eschews any form of the skeptic's concerns simply by denying any need for further justification of the reliable process \textit{m}. In fact, it is a well known characteristic of process reliabilism that it rejects any form of regression in the justificatory chain. That is, whether or not we know that the method by which we attribute reliability is, in and by itself, reliable, is of no concern to the traditional reliabilist \cite{Goldman2016, Bird1998}. In this sense, and always according to process reliabilism, Peter is justified in believing that `2 + 2 = 4' because he acquired true beliefs by looking into a textbook on algebra, and algebra is a reliable process as a matter of fact. There is no need for further justification that algebra is, in turn, a reliable process. 

In its standard form, process reliabilism is inapplicable as a general solution to EEO without providing further restrictions. Specifically, process reliabilism do not require the agent to know (or to justify) the methods which produce reliable processes. This is the so called JJ-principle, which states that in order for a method to yield justified belief, the method too must be justified \cite[152]{Bird1998}. By evading this principle, process reliablism is unable to account for the varied justificatory practices as detailed in our paper. Why scientists concern themselves with verification and validation, robustness analysis, etc. and how the epistemic strength of those methods is to be evaluated are questions which cannot be answered by process reliabilism. In computational reliablism, instead, an agent can know something by relying on a reliable method (i.e., the computational process). In this first step, the JJ-principle is not required, thereby circumventing the skeptical challenge posed by Humphreys. But as we would also like to allow skepticism about reliable methods, we reintroduce the JJ-principle in the second step. If a researcher thinks that a specific verification procedure is reliable, she would have to adduce reasons for it. In this way, computational reliabilism better reflects current scientific and engineering practices better.

Thus understood, computational reliabilism requires a `retrospective reliability chain,' one that conditions the sources that attribute reliability to computer simulations to be reliable in and by themselves. This means that the sources presented in section \ref{four_sources} must be shown to be reliable. For instance, many verification and validation methods depend, in turn, on mathematics and the empirical sciences. The history of (un)successful implementations, on the other hand, is a reliable source insofar as there are well defined theories in the social epistemology and scientific practice that can vouch for the methods that populate such history. As more sources come into play, or as the same sources change over time, researchers must sanction their reliability. 


To sum up, computational reliabilism encompasses two specifications of standard process reliabilism and one amendment. These are, the truth-valued proposition \textit{p}, which stands for the results of the computer simulation; the reliable computer simulation \textit{m}, which is a specification of the reliable process; and the series of reliable sources leading to the reliability of \textit{m}.


\section{Sources for computational reliabilism} \label{four_sources}

In the following, we identify four sources for attributing reliability to a computational process such as computer simulations. It is important to note that each source offers a different `degree of reliability' to computer simulations. For instance, expert knowledge by itself is a rather weak source for the reliability of most computer simulation. The reason for this is that it could be idiosyncratic in several ways, and therefore not reliable in the epistemic sense required. Verification and validation methods, on the other hand, are stronger forms of reliability for they depend on mathematical machinery and thus are epistemically more secure. This is the reason why the latter, and not expertise knowledge, are on many occasions decisive for attributing reliability to computer simulations. Having said this, we are unable to offer here a measurement of the degree of reliability for each source. Instead, we offer an analysis of each individual source.


\begin{enumerate}
	\item Verification and validation methods
	\item Robustness analysis for computer simulations
	\item A history of (un)successful implementations
	\item Expert knowledge
\end{enumerate}

\subsection{Verification and validation}

\emph{Verification} and \emph{validation}\footnote{Also known as `internal validity' and `external validity' or `testing' respectively.} are the general names given to a host of methods used for increasing the reliability of scientific models as well as computer simulations. Understanding their role, then, turns out to be essential for attributing reliability to computer simulations. 

In \emph{verification}, it is standard that formal methods are at the center for the reliability of computer software, whereas in \textit{validation} benchmarking is responsible for confirmation of the outcomes \cite[Preface]{Oberkampf2010}. In verification methods, then, the relationship of interest is between the specification of a model and the computer software, whereas in validation methods the relationship of interest is between computation and the empirical world. Here are two standard definitions largely accepted and used by the community of researchers:

\begin{quote}
	\noindent \emph{Verification}: the process of determining that a computational model accurately represents the underlying mathematical model and its solution.
	
	\noindent \emph{Validation}: the process of determining the degree to which a model is an accurate representation of the real world from the perspective of the intended uses of the model. \cite{Oberkampf2003}
\end{quote} 

In recent philosophical studies, these definitions have been adapted to include computer simulations. Eric Winsberg, for instance, takes it that ``\emph{verification}, [...] is the process of determining whether or not the output of the simulation approximates the true solutions to the differential equations of the original model. \emph{Validation}, on the other hand, is the process of determining whether or not the chosen model is a good representation of the real-world system for the purpose of the simulation'' \cite[19-20]{Winsberg2010}. Another example of a philosopher discussing verification and validation in computer simulations is Margaret Morrison. Although she agrees with Winsberg that verification and validation are two methods not always clearly divisible, she nevertheless downplays the need for verification methods claiming that validation is a more crucial method for assessing the reliability of computer simulation \cite[43]{Morrison2009}.

The scientific and computational communities, in contrast, have a more diverse set of definitions to offer, all tailored to the specificities of the simulation under study. In \textit{verification} studies, for instance, the literature provides two methods particularly important for computer simulations. These are \textit{code verification} and \textit{calculation verification}.\footnote{Also referred to as \emph{solution verification} in \cite[26]{Oberkampf2010}, and as \emph{numerical error estimation} in \cite[26]{Oberkampf2003}.} Their importance lies in the fact that both methods focus on the correctness of the discretization procedure, a key element for implementing mathematical models as computer simulations.


William Oberkampf and Timothy Trucano have further argued that it is useful to segregate code verification into two activities, namely, \textit{numerical algorithm verification} and \textit{software quality engineering}. The purpose of numerical algorithm verification is to address the mathematical correctness of the implementation of all the numerical algorithms that affect the numerical accuracy of the results of the simulation. The goal of this verification method is to demonstrate that the numerical algorithms implemented as part of the simulation model are correctly implemented and performing as intended \cite[720]{Oberkampf2002}. Software quality engineering, on the other hand, sets the emphasis on determining whether the simulation model is reliable and produces, most of the time, trustworthy results. The purpose of software quality engineering is to verify the simulation model and the results of the simulation on a specific computer hardware, in a specified software environment -- including compilers, libraries, I/O, etc. These verification procedures are primarily in use during the development, testing, and maintenance of the simulation model \cite[721]{Oberkampf2002}. 

As for \emph{calculation verification}, it is generally depicted as the method that prevents three kinds of errors: human error in the preparation of the code, human error in the analysis of the results, and numerical errors resulting from computing the discretized solution of the simulation model. A definition for calculation verification is ``the process of determining the correctness of the input data, the numerical accuracy of the solution obtained, and the correctness of the output data for a particular simulation'' \cite[34]{Oberkampf2003}.


The process of \emph{validation} consists in showing that the results of the simulation correspond, more or less accurately and precisely, to those obtained by measurement and observation of the target system. Oberkampf and Trucano highlight three key aspects of validation methods. These are ``i) quantification of the accuracy of the computational model by comparing its responses with experimentally measured responses, ii) interpolation or extrapolation of the computational model to conditions corresponding to the intended use of the model, and iii) determination if the estimated accuracy of the computational model, for the conditions of the intended use, satisfies the accuracy requirements specified'' \cite[724]{Oberkampf2008}.

It is important to mention that, with the introduction of computer simulations in experimental contexts, validation does not exclusively depend on contrasting results against empirical data. Ajelli and team have shown how it is possible to run different computer simulations and use their results to assert their mutual reliability -- in this case, there is not a mere convergence of results, but also of key variables \cite{Ajelli2010}, as we argue in section \ref{Robustness}.

The role of verification and validation methods in attributing reliability to computer simulations is rather straightforward: on the one hand, they make sure that the implementation of well established theories is correctly carried out and not much information is missed; on the other, they provide good reasons to trust the results of the simulations because they match, with more or less accuracy, empirical data.


\subsection{Robustness analysis for computer simulations} \label{Robustness}

When systems under study are inherently too complex and particular degrees of precision and accuracy in idealized models are required but not delivered by fundamental theories, then \textit{robustness analysis} becomes a suitable alternative method for determining the trustworthiness of results \cite[156]{Weisberg2013}. 

Robustness analysis, as presented by Richard \cite{Levins1966} and further elaborated by Michael \cite{Weisberg2013} allows researchers to learn about the results of a given model and whether they are an artifact of it (e.g., due to a poor idealization) or whether they are related to core features of the model \cite[156]{Weisberg2013}. At its heart, robustness analysis consists of two steps, the first one consisting in examining a group of models to determine if they all predict a common result -- called the \textit{robust property}; during the second step, models are analyzed for those structures in the model that generate the sought robust property. The results from these two steps are combined in order to formulate the \textit{robust theorem}, ``a conditional statement linking common structure to robust property, prefaced by a \textit{ceteris paribus} clause'' \cite[158]{Weisberg2013}. It is important to emphasize that robust theorems do not make claims about the frequency with which the robust property occur in target systems. Rather, it makes the conditional claim about what happens if a model is instantiated in an specific way \cite[169]{Weisberg2013}.

Following Weisberg, the ideal case of robustness analysis requires researchers to examine a group of similar but distinct models in search of a robust behavior. The aim of such an examination is to formulate sufficiently diverse models in such a way that the discovery of a robust property is not due to mere luck in the way the models were analyzed but rather because the property is actually there \cite[158]{Weisberg2013}. The question now is how to formulate such diverse models. Weisberg suggests a list of possibilities, none of which consists of changes in the parametrization of the model and of initial and boundary conditions, but in significant modifications to the structure of the model. Reinterpreting these possibilities in terms of modification in computer simulations, they include varying the regularity of the grid, varying the number of attributes of a process, and varying the heterogeneity of the utility function, among others.

Let us note that Weisberg's analysis of robustness relies on the number of (heterogeneous) models that researchers are able to create. The more models available, the more likely it is that the robust property identified across models can actually be found in a real-world system \cite[160ff]{Weisberg2013}. In computer simulations, the computational power allows researchers to produce a large number of heterogeneous models at a relatively low cost (e.g., in terms of human resources, money, time, etc.). In this sense, inferring that a robust property is present in the simulation models, and therefore that the core structure is giving rise to such a property, is a much simpler task with computer simulations. 

Now, the core assumption in robustness analysis is that if a sufficiently heterogeneous set of models give rise to a property, then it is very likely that the real-world phenomenon also shows the same property. Furthermore, robustness analysis allows researchers to infer that, when the robust property is observed in a real-world system, then it is very likely that the core structure of the computer simulation corresponds to the causal structure giving rise to the real-world phenomenon. Robustness analysis, therefore, is a key player in the process of attributing reliability to computer simulations. 

Consider the following example of robustness analysis in computer simulations. Ajelli et al. provide a side-by-side comparison of two computer simulations, a stochastic agent-based model and a structured meta-population stochastic model (GLobal Epidemic and Mobility - GLEaM). The agent-based model includes an explicit representation of the Italian population through highly detailed data on the socio-demographic structure. In addition, and for determining the probability of commuting from municipality to municipality, Ajelli et al. use a general gravity model used in transportation theory. However, the epidemic transmission dynamics is based on an ILI (Influenza-like Illness) compartmentalization, which in turn is based on stochastic models that integrate susceptible, latent, asymptomatic infections, and symptomatic infections \cite[5]{Ajelli2010}. The authors define their agent-based model as ``a stochastic, spatially-explicit, discrete-time, simulation model where the agents represent human individuals [...]  One of the key features of the model is the characterization of the network of contacts among individuals based on a realistic model of the socio-demographic structure of the Italian population.'' \cite[4]{Ajelli2010} The authors also mention that both GLEaM and the agent-based model are dynamically calibrated in that they share exactly the same initial and boundary conditions \cite[6]{Ajelli2010}.

On the other hand, GLEaM is a multiscale mobility network based on high-resolution population data that estimates the population with a resolution given by cells of 15  x 15 minutes of arc. Balcan et al. explain that a typical GLEaM consists of three data layers. A first layer, where the population and mobility allows the partition of the world into geographical regions. This partition defines a second layer, the subpopulation network, where the inter-connection represents the fluxes of individuals via transportation infrastructures and general mobility patterns. Finally, and superimposed onto this layer, is the epidemic layer, that defines inside each subpopulation the disease dynamic \cite{Balcan2009}. In the study by Ajelli et al., GLEaM also represents a grid-like partition where each cell is assigned the closest airport. The subpopulation network uses geographic census data, and the mobility layers obtain data from different databases, including the International Air Transport Association database consisting in a list of airports worldwide connected by direct flights.

By increasing spatial resolution, changing grid size, the topography of the network, internal functions, and several other structures -- tailored to what each model can offer to alter -- Ajelli et al. are able to identify a series of robust properties and thus elaborate a series of robust theorems.\footnote{A further point to evaluate is whether identifying differences in what should be a robust property is epistemically as relevant as identifying a robust property. The former requires an evaluation by the researchers of what \textit{should} be a robust property whereas the later is somehow provided by the simulation model. To Ajelli et al. identifying similarities and differences both work towards the reliability of their computer simulation: ``we investigated and quantified similarities and differences in the results at different scales of resolution, and related those to the assumptions of the frameworks and to their integrated data'' \cite[11]{Ajelli2010}} To illustrate just one case, Ajelli et al. reported to have found that the two computer simulations ``display a very good agreement in the timing of the epidemic, with a very limited variation in the time of the simulated epidemic activity peaks. In the metapopulation approach the fraction of the population affected by the epidemic is larger (by 5\% to 10\%) than in the agent-based approach. This difference is due to the assumption of homogeneity and thus the lack of detailed structure of contacts (besides the age structure) in the metapopulation approach with respect to the agent-based approach''  \cite[11]{Ajelli2010}. In this case, robustness analysis provides good reasons to believe that core structures in GLEaM and the agent-base simulation correspond very well to the actual timing of the epidemic. Researchers are thus justified in believing claims about results of these simulations -- and from those created from these two simulations.

\subsection{A history of (un)successful implementations}


The history of science offers a long record of successes and accomplishments, as well as failures and incompetence. What does such a disruptive history tell us about the scientific enterprise? In the context of experimental practice, Ian \cite{Hacking1988a} and Peter \cite{Galison1997} have argued that mature science has been, by and large, cumulative since the seventeenth century. Such a claim builds on the idea that (un)successful implementation of a theory, a model, or even two chemicals in a laboratory setup are part of the corpus of knowledge as much as the theory, the model, and the two chemicals in question.

Something very similar can be said about the success, failure and cumulative nature of computer simulations. The simulation model as a whole is conceptualized, designed, programmed and executed in a series of stages that do not remain constant over time \cite{Duran2018}. In each stage, the knowledge relied upon to devise each method comes from a wide range of domains, including mathematics, logic and computer theory, sociology and cognitive psychology. Over time, techniques are improved upon, reconfigured, and radically revised when the technology changes or a new method is envisaged. For instance, \textit{design prototyping} is a sub-field of software engineering that helps developers assess alternative design strategies and decide which is best for a particular project. There are no standard methods for choosing the best strategy, but rather the designers may address the requirements of the simulation with several different design approaches to see which has the best properties. For instance, a simulation involving networking may be built as a ring in one prototype and as a star in another, and performance characteristics evaluated to see which structure is better at meeting performance goals or constraints \cite[Chapter 5]{Pfleeger2009}. In this respect, for some cases the best option will be to draw from a body of successful implementations (e.g., of successful implementations of ring networking simulations); for some other cases, a new strategy will populate such a body (e.g., failures in communication protocols, and the success in a new networking topology). In both cases, they integrate a history of (un)successful implementations. 

This is, we believe, part of what \cite{Massimi2015} have in mind when they claim that the epistemic reliability of computer simulations come from the credentials supplied by well grounded scientific knowledge. Although we agree with this claim, we must keep in mind that the methodology of computer simulations is dynamic and non-hierarchical. That is to say that researchers make constant changes to their simulations, rather than merely implementing a well grounded theory once and for all. It is also to say that well grounded scientific knowledge is, to today's scientific standards, also knowledge generated by computer simulations. In this vein, well grounded scientific knowledge depends as much on computer simulations as the latter depend on scientific knowledge. Naturally, such a dynamism in the methodology might introduce sources of unreliability (e.g., using a method that has been historically successful in one domain into a complete different domain). However, the simulation model itself is, at some point, methodologically stabilized -- as opposed to constant tinkering.

In this respect, we follow Eric Winsberg who, borrowing in turn from Hacking, claimed that building techniques have their own life for ``they carry with them their own history of prior (un)successes and accomplishments, and, when properly used, they can bring to the table independent warrant for belief in the models they are used to build'' \cite[122]{Winsberg2003}. We include such history of (un)successful implementations as an important source for attributing reliability to computer simulations. 

\subsection{Expert knowledge}

The last source we offer here for computer reliabilism can be found in the different disciplines that constitute Science and Technology Studies. In there, a great deal of attention is put on understanding the notion and role of experts in science and engineering. Harry Collins and Robert Evans argue that standard theories of expertise (e.g., the \textit{relational theory of expertise}, which take expertise to be a matter of the experts' relations with other experts \cite[2]{Collins2007}) fall short in a series of respects. They usually provide no guidance on how to legitimize and identify the experts nor how to choose between competing experts (see the \textit{periodic table of expertises} \cite[14]{Collins2007}); furthermore they leave out of consideration the analysis of the citizen's role in technological decision-making and, if the proper measures are not in place, they can be dangerously idiosyncratic. Collins and Evans propose as alternative the \textit{realist theory}, which takes that expertise is some sort of attribute or possession that groups of experts have and that individuals acquire through their membership of those groups. ``Acquiring expertise'' Collin and Evans conclude, ``is therefore a social process -- a matter of socialization into the practices of an expert group -- and expertise can be lost if time is spent away from the group'' \cite[3]{Collins2007}.

To us, the expert is interpreted in the realist mode proposed by Collin and Evans, with the condition that having membership of a given group does not mean strict participation in that group. Thus, to us the mathematician and physicist that know the underlying theory that will be implemented as a simulation very well, but know nothing about the implementation itself, are as much an expert in the computer simulation as the computer scientist that knows how to implement the theory but little or nothing about the theory itself. 


As Claus Beisbart indicates, scientists believe the results of their simulations because they trust the assumptions upon which such simulations are built \cite{Beisbart2017}. These assumptions are here interpreted as being suggested and approved by the relevant actors, that is, the experts. Furthermore, by and large scientists believe the results of their simulations because they fall within an expected range. Marco Ajelli et al. provide us with a good example of the interplay between the assumptions built into the simulation model and what experts typically anticipate. To Ajelli et al. ``[t]he epidemic size profile shows an expected overall mismatch of 5-10\% depending on the reproductive rate, which is induced by the homogeneous assumption of the metapopulation strategy'' \cite[2]{Ajelli2010}. 

With these ideas in mind, it is possible to argue that the expert is a key contributor to the reliability of computer simulations:\footnote{Let us keep in mind that expert knowledge is also a source for a variety of errors. Thanks to an anonymous reviewer for pointing this out to us.} the theory and assumptions built into the simulation, along with the implicit theory supporting the computation largely depend on the experts, and/or they determine the range within which results can be accepted. 

Expert knowledge also plays an important role in determining the robustness of a simulation as well as in participating in a history of (un)successful implementations. In the latter case, because they are the main actors in creating such (un)successful history. In the former case, because the expert's abilities to identify and judge relevantly similar structures is paramount for claims about robust properties. According to Weisberg, there are occasions where researchers rely on judgment and experience, not mathematics or simulation, to determine whether a common structure gives rise to the robust behavior as well as judge whether the common structure contains important mathematical similarities as opposed to just intuitive qualitative similarities \cite[159]{Weisberg2013}. Ajelli et al. again offer an interesting assertion that combines claims about robustness and the modeling assumptions advanced by experts: ``[t]he good agreement of the two approaches [i.e., the agent-based simulation and the GLEaM simulation] reinforces the message that computational approaches are stable with respect to different data integration strategies and modeling assumption'' \cite[2]{Ajelli2010}

\section{Final remarks} \label{final_remarks}

If the philosophical novelty of computer simulations is a matter of controversy, we hope that this article evidenced that more philosophical efforts need to be channeled towards a better understanding of simulations. The EEO is just one issue proposed for a genuine philosophy of computer simulations. To the few attempts to answer it, we suggested a peaceful coexistence between accepting EEO and reasons for having genuine knowledge provided by computer simulations. 

Although our approach builds much from past research on justification and theories of knowledge, it is new in at least three different ways. First, because it is the only account that takes EEO seriously and proposes an effective solution to it in terms of theories of knowledge; second, because it attends to some shortcomings that process reliabilism has in the context of computer simulations (hence, renaming it \textit{computational reliabilism}); and third, because it is the only work in the literature that systematically and qualitatively addresses the sources that attribute reliability to computer simulations. 

Having mentioned these merits, we must also acknowledge the limitations of our approach. For starters, more needs to be said about the sources attributing computational reliability. Whereas our treatment has been general, some specific work would shed more light on the methods for attributing reliability to computer simulations. For instance, addressing verification methods exclusively for computer simulations will help to understand better the degree to which they are absolutely necessary in the assessment of their reliability. Similarly, to the argument here advanced, it is important to factor in the specific uses of computer simulations. In climate science, for instance, expert knowledge has a more epistemically prominent place than validation methods because the scarcity of data makes the justification of the simulation \textit{via} validation a rather difficult task. This is, of course, not to say that we must accept weaker standards of verification and validation for climate science. Rather that the justification of certain simulations, such as in climate science, comes by and large from expertise knowledge.\footnote{We thank an anonymous reviewer for helping us clarify this point.}

All in all, we expect to have provided a formal account of how to address and solve the skeptic's challenge that follows by taking EEO seriously in the context of computer simulations.

	\bibliographystyle{plain}
	\bibliography{./EEO.bib}
	
\end{document}